\documentclass[%
 aip,
 jap,%
 amsmath,amssymb,
 reprint,%
]{revtex4-1}

\usepackage{graphicx}
\usepackage{dcolumn}
\usepackage{bm}
\begin{document}

\title{Magnetic and magnetocaloric properties of quasi-one-dimensional Ising spin chain CoV$_{2}$O$_{6}$}

\author{M. Nandi and P. Mandal}
\email{prabhat.mandal@saha.ac.in}
\affiliation{Saha Institute of Nuclear Physics, 1/AF Bidhannagar, Calcutta 700 064, India}
\date{\today}

\begin{abstract}

We have investigated the magnetic and magnetocaloric properties of  antiferromagnetic Ising spin chain CoV$_{2}$O$_{6}$ by magnetization and heat capacity measurements. Both monoclinic $\alpha$-CoV$_{2}$O$_{6}$ and triclinic $\gamma$-CoV$_{2}$O$_{6}$ exhibit field-induced metamagnetic transition from antiferromagnetic  to ferromagnetic  state via an intermediate ferrimagnetic state with 1/3 magnetization plateau.  Due to this field-induced metamagnetic transition, these systems show large conventional as well as inverse magnetocaloric effects. In $\alpha$-CoV$_{2}$O$_{6}$, we observe field-induced complex magnetic phases and multiple magnetization plateaux at low temperature  when the field is applied along $c$ axis. Several critical temperatures and fields have been identified from the temperature and field dependence of magnetization, magnetic entropy change and heat capacity to construct the $H$-$T$ phase diagram.  As compared to $\alpha$-CoV$_{2}$O$_{6}$, $\gamma$-CoV$_{2}$O$_{6}$ displays a relatively simple magnetic phase diagram.  Due to the large magnetic entropy  change and adiabatic temperature change at low or moderate applied magnetic field, $\gamma$-CoV$_{2}$O$_{6}$ may be considered as a  magnetic refrigerant in the low-temperature region.
\end{abstract}


\maketitle

\section{Introduction}

Combination of low dimensionality and geometrical frustration sometimes reveal complex magnetic behavior at low temperature in several magnetic systems. For example, quasi-one-dimensional Ising spin chain CsCoCl$_{3}$ exhibits magnetic transitions around 33  and 44 T due to the complex intrachain and interchain spin rearrangements.\cite{amaya} Another spin system Ca$_{3}$Co$_{2}$O$_{6}$ having a triangular lattice arrangement of ferromagnetic (FM) Ising chains coupled via weak antiferromagnetic (AFM) exchange has become a center of both theoretical and experimental study due to its peculiar behavior of magnetic properties at low temperature\cite{maignan,soto,hardy,maignan1,kudasov} and a complex magnetic phase diagram has been constructed from the field and temperature dependence of magnetic entropy change. \cite{lam} Several studies have also demonstrated that the field-induced metamagnetic transition often exhibits large magnetic entropy change which may be suitable for magnetic refrigeration. Cooling based on magnetocaloric effect (MCE) is one of the remarkable discoveries in the field of refrigeration technology because it does not use the harmful chlorofluorocarbon gas.\cite{kag} Large isothermal magnetic entropy change ($\Delta$S$_{M}$) and adiabatic temperature change ($\Delta T_{ad}$) are the two important parameters that take decisive role on the technological applicability of a material in  magnetic refrigeration. Scientists and engineers are looking for  materials having large MCE at low field near room temperature for domestic and industrial refrigerant purposes. However, large MCE at low temperature  can be used for space science application and liquefaction of hydrogen in fuel industry. Till now, magnetic refrigeration in the low temperature region ($<$20 K) is done by using paramagnetic (PM)  salt.\\

In recent years, quasi-one-dimensional Ising spin chain CoV$_{2}$O$_{6}$ has attracted  interest in  research  due to its interesting properties like field-induced metamagnetic transition with 1/3 magnetization plateau, large single ion anisotropy, strong spin-lattice coupling, etc. CoV$_{2}$O$_{6}$ shows two types of crystalline phases depending on the local environment.\cite{lene,kim,holl} One is high temperature monoclinic $\alpha$-CoV$_{2}$O$_{6}$ with C2/m space group \cite{zhe,lene,kim,holl,lener,ksingh,mark,markk,nandi} and the another is low temperature triclinic $\gamma$-CoV$_{2}$O$_{6}$ with P-1 space group.\cite{zh,lene,kim,kimb,holl} Both $\alpha$-CoV$_{2}$O$_{6}$ and $\gamma$-CoV$_{2}$O$_{6}$ are constituted of edge-shared CoO$_6$ octahedra forming 1D spin chain along $b$ axis and the edge-shared VO$_5$ square-pyramids are located in between the spin chains. Though spin chains are similar in nature for both the phases, the direction of their easy axis of magnetization is different.  The easy axis is along the $c$ axis in $\alpha$ phase \cite{zhe,lener} while it is along the chain direction  in  $\gamma$ phase.\cite{lenert,lenertz} As V ions are in nonmagnetic pentavalent state, the magnetic contribution comes from the Co$^{2+}$ ion. The intrachain magnetic interaction is strong and ferromagnetic  in nature while the chains are weakly coupled via antiferromagnetic  interaction. Both $\alpha$ and $\gamma$ phases show AFM transition with Neel temperatures ($T_N$) 14 K and 6.5 K, respectively.\cite{zhe,lene,kimb}\\

In this work, we have compared and contrasted the magnetic and magnetocaloric properties of magnetically aligned $\alpha$-CoV$_2$O$_6$ and $\gamma$-CoV$_{2}$O$_{6}$. The present study reveals several important differences between field-induced magnetic states in these two systems and demonstrates that  $\gamma$-CoV$_{2}$O$_{6}$ is suitable for  magnetic refrigerantion in the low-temperature region  due to its larger magnetic entropy  change and adiabatic temperature change at low or moderate applied magnetic field.  The detailed investigation on the nature of multiple magnetic transitions at low temperature points out some interesting behavior which needs further theoretical and microscopic investigation for the $\alpha$-CoV$_{2}$O$_{6}$ system. \\

\section{Experimental Details}
\begin{figure}[b!]
\begin{center}
\includegraphics[width=0.50\textwidth]{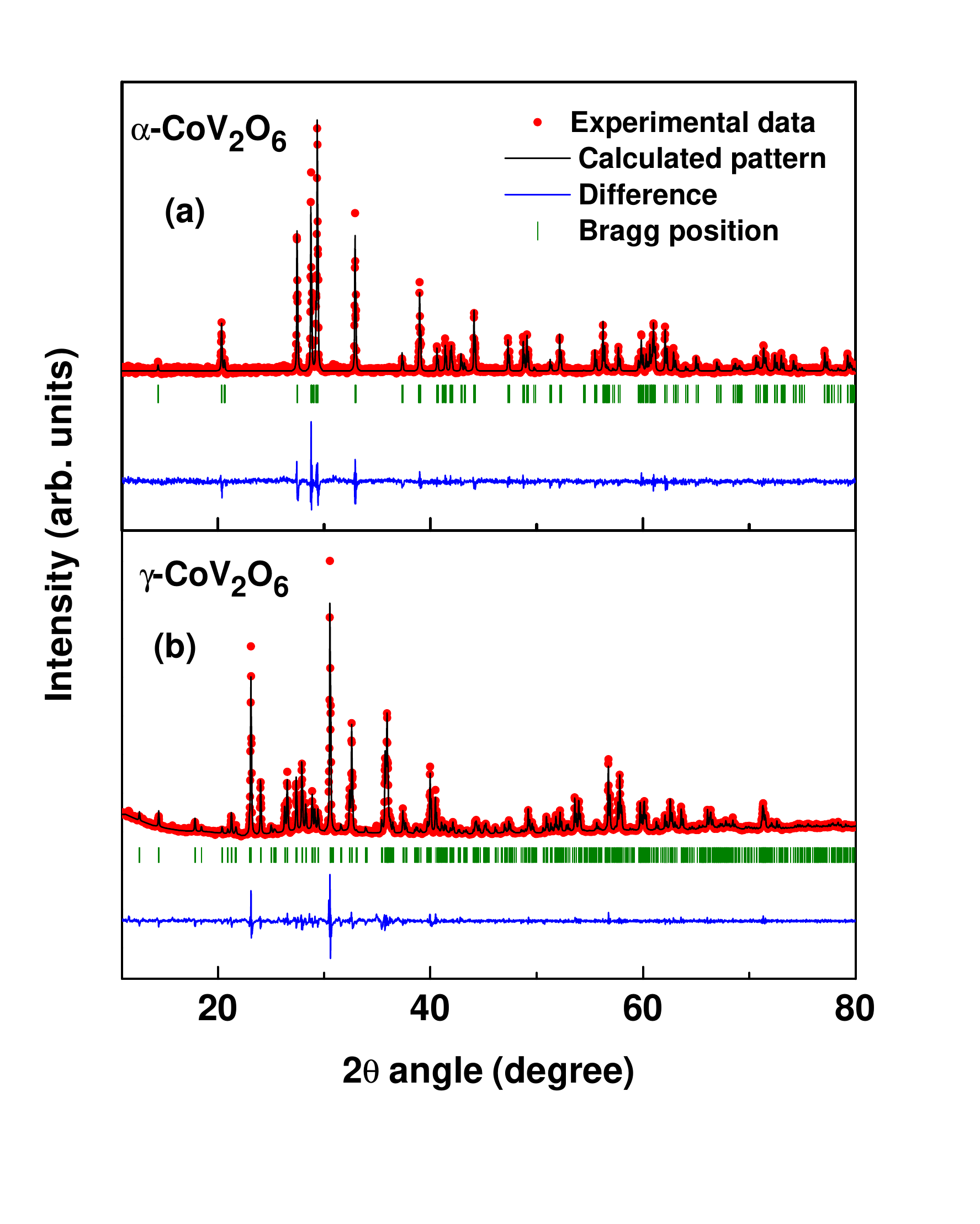}
\caption{X-ray powder diffraction patterns at room temperature for $\alpha$-CoV$_{2}$O$_{6}$ (a) and $\gamma$-CoV$_{2}$O$_{6}$ (b). The black solid line corresponds to the Rietveld refinements of the diffraction pattern.}
\end{center}
\end{figure}
Polycrystalline  $\alpha$-CoV$_{2}$O$_{6}$ and $\gamma$-CoV$_{2}$O$_{6}$ samples were prepared by standard solid-state reaction method using the mixture of stoichiometric quantities of high purity V$_2$O$_5$ and cobalt acetate tetrahydrate or cobalt oxalate dihydrate. For $\alpha$-CoV$_{2}$O$_{6}$,  the mixture was heated in air at 650$^{\circ}$C for 16 h and then at 725$^{\circ}$C for 48 h. After the heat treatment, the material was quenched in liquid nitrogen  to obtain single phase $\alpha$-CoV$_{2}$O$_{6}$. For $\gamma$-CoV$_{2}$O$_{6}$, the mixture was heated in air at 620$^{\circ}$C for 45 h and then cooled to room temperature at a rate of 2$^{\circ}$C/min. The resulting powder was ground properly and pressed into pellets. Finally, the pellets were heated again at 620$^{\circ}$C  for 45 h and cooled at a rate of 2$^{\circ}$C/min. Phase purity of these compounds was checked by powder x-ray diffraction (XRD) method with CuK$_{\alpha}$ radiation in a  Rigaku TTRAX II  diffractometer. Figure 1 shows the X-ray powder diffraction pattern for $\alpha$-CoV$_{2}$O$_{6}$ and  $\gamma$-CoV$_{2}$O$_{6}$ samples at room temperature. No trace of impurity phase was detected within the resolution of XRD. All the peaks in XRD were assigned to a monoclinic structure of space group $C$2/$m$ for $\alpha$ phase whereas to a triclinic structure of space group P-1 for $\gamma$ phase using the Rietveld method. The observed lattice parameters  \emph{a}=9.2501 {\AA}, \emph{b}=3.5029 {\AA}, \emph{c}=6.6175 {\AA} and $\beta$=111.61$^{\circ}$  for $\alpha$ phase and \emph{a}=7.180 {\AA}, \emph{b}=8.899 {\AA}, \emph{c}=4.812 {\AA}, $\alpha$= 90.28$^{\circ}$, $\beta$=93.66$^{\circ}$  and $\gamma$=102.21$^{\circ}$ for $\gamma$ phase are in good agreement with the reported values. \cite{lene,zhe,lener,ksingh,mark} For the magnetic alignment, the polycrystalline powder was mixed with a polymeric gel and  kept at room temperature for 20 h in presence of 7 T magnetic field. Huge magnetic anisotropy helps the microcrystals in powder to align their easy axis of magnetization along the magnetic field direction and makes the system equivalent to single crystal with field parallel to crystallographic  $c$ and $b$ axes for $\alpha$-CoV$_{2}$O$_{6}$ and $\gamma$-CoV$_{2}$O$_{6}$, respectively. The magnetization was measured using a SQUID-VSM (Quantum Design) with field along the easy axis. The  zero-field specific heat ($C_p$) measurements were done on both random and aligned polycrystalline samples using a physical property measurement system (Quantum Design). \\

\section{Results and discussion}
\begin{figure}
\begin{center}
\includegraphics[width=0.50\textwidth]{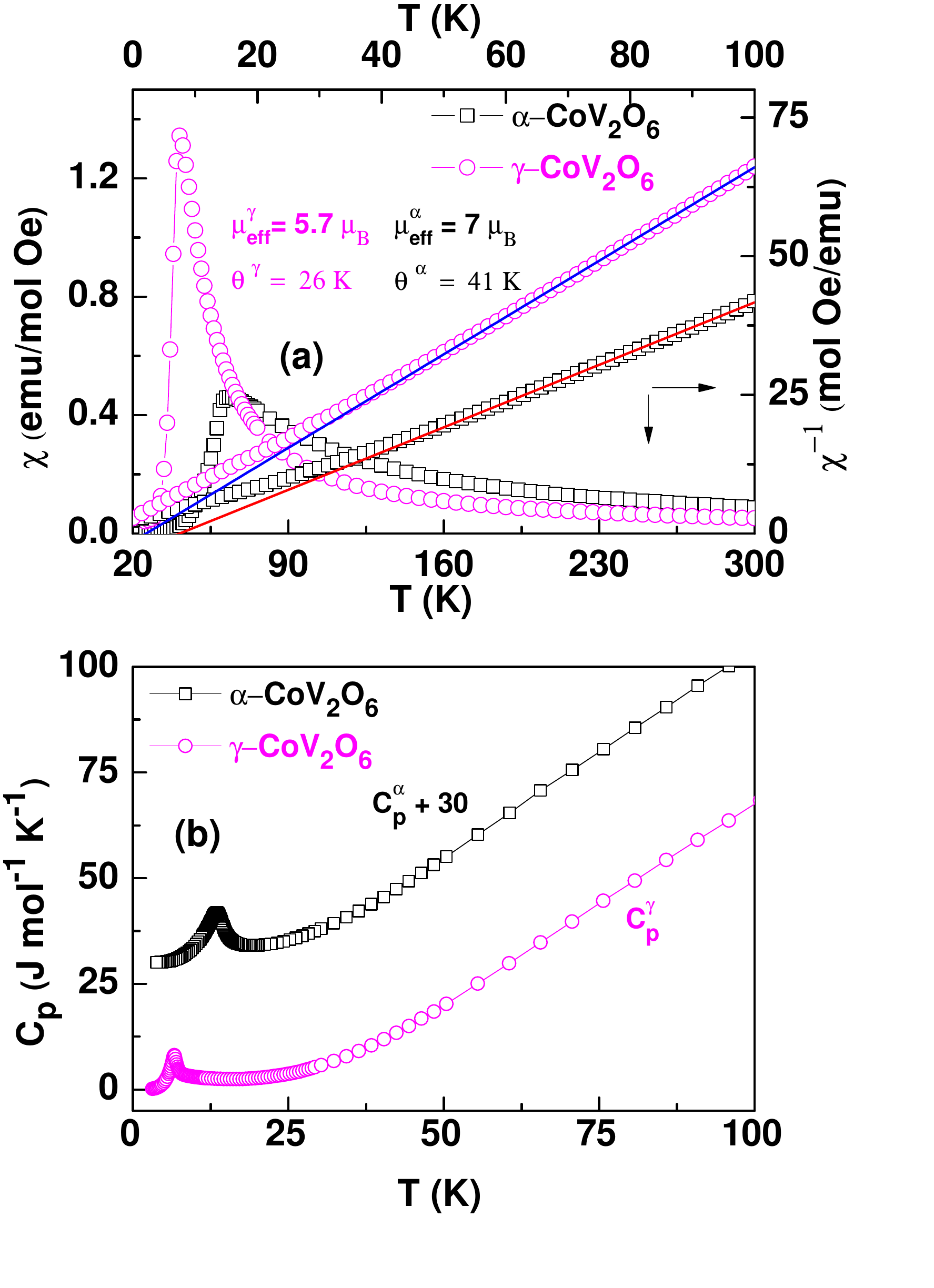}
\caption{(a) Temperature dependence zero-field-cool susceptibility ($\chi$) for $\alpha$-CoV$_{2}$O$_{6}$ at 1000 Oe and $\gamma$-CoV$_{2}$O$_{6}$ at 500 Oe. The right axis shows the inverse of susceptibility  ($\chi$$^{-1}$)  and the corresponding Curie-Weiss fit (solid line). (b) Zero-field heat capacity as a function of temperature for random $\alpha$-CoV$_{2}$O$_{6}$ and $\gamma$-CoV$_{2}$O$_{6}$. For clarity,  the heat capacity data for $\alpha$-CoV$_{2}$O$_{6}$ have been shifted upwards by 30 units along the $y$ axis.}
\end{center}
\end{figure}
The zero-field-cool susceptibility ($\chi$) and inverse susceptibility ($\chi^{-1}$) versus temperature for $\alpha$-CoV$_{2}$O$_{6}$ and $\gamma$-CoV$_{2}$O$_{6}$  compounds are presented in Fig. 2(a). Both $\alpha$-CoV$_{2}$O$_{6}$ and $\gamma$-CoV$_{2}$O$_{6}$  exhibit a sharp peak due to the AFM transition  around $T_N$$=$14 K and 6.5 K, respectively.  $\chi^{-1}$($T$) can be fitted well with the Curie-Weiss law [$\chi$$=N\mu_{eff}^2/3k_B(T-\theta$)] over a wide range of temperature (200-300 K) well above $T_N$ as shown in Fig. 2(a). From this fitting,  we have deduced effective moment $\mu_{eff}^{\alpha}$$=$7 $\mu_B$/Co ion and the Weiss temperature $\theta^{\alpha}$$=$41 K  for $\alpha$-CoV$_{2}$O$_{6}$ whereas  $\mu_{eff}^{\gamma}$$=$5.7 $\mu_B$/Co ion and $\theta^{\gamma}$$=$ 26 K  for $\gamma$-CoV$_{2}$O$_{6}$.  The positive value of $\theta$ is reflecting the strong intrachain FM interaction in these systems. The above  values of $\mu_{eff}$ and $\theta$ are comparable with that reported for single crystals. \cite{zhe,lenertz}  The observed values of effective moment are much larger than the expected spin-only moment of high spin Co$^{2+}$ (3.87 $\mu_B$). This suggests that one should consider the spin-orbit coupling to compare the deduced values of magnetic moment with theoretical ones.  We discuss the role of spin-orbit coupling in details in the later section.  The temperature dependence of zero-field specific heat for the polycrystalline random samples also confirms AFM transition around 14 K and 6.5 K for $\alpha$-CoV$_{2}$O$_{6}$ and $\gamma$-CoV$_{2}$O$_{6}$, respectively [Fig. 2(b)]. For both the systems, the $\lambda$-like anomaly in specific heat suggests that the transition is second order in nature.\\
\begin{figure}
\begin{center}
\includegraphics[width=0.50\textwidth]{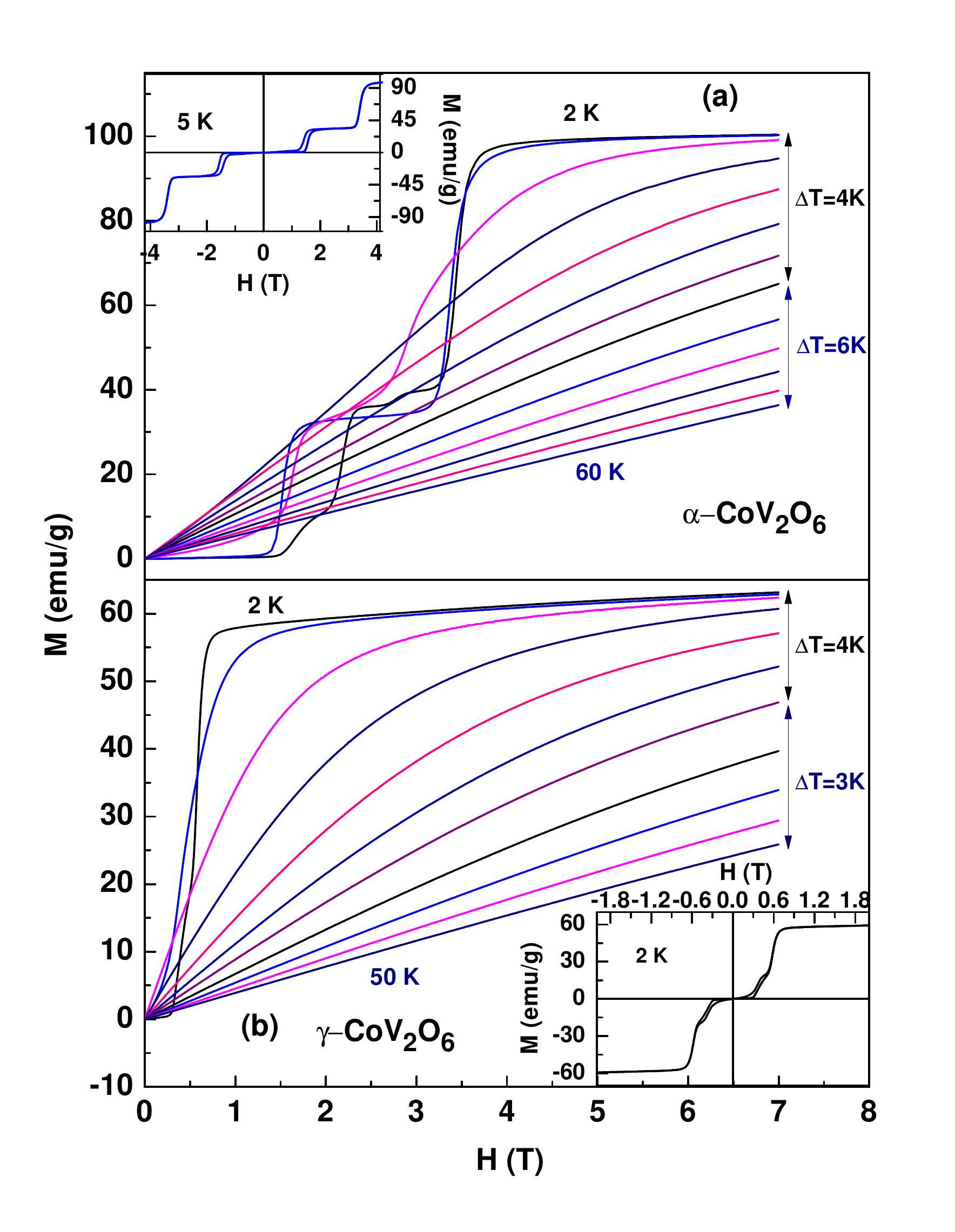}
\caption{(a) and (b) Isothermal magnetization for $\alpha$-CoV$_{2}$O$_{6}$ and $\gamma$-CoV$_{2}$O$_{6}$. Insets show the corresponding low-field hysteresis at 5 K for $\alpha$-CoV$_{2}$O$_{6}$ and at 2 K for $\gamma$-CoV$_{2}$O$_{6}$.}
\end{center}
\end{figure}
The isothermal magnetization curves  are displayed in Figs. 3(a) and (b). For the sake of clarity, the data with increasing field for some selected temperatures are shown. Well below $T_N$, the $M$($H$) curve  exhibits  step-like increase at two critical fields $H_{c1}$ and  $H_{c2}$. For $\alpha$-CoV$_{2}$O$_{6}$, $H_{c1}$ and  $H_{c2}$ are 1.6 and 3.3 T, respectively while the corresponding values are 0.36 and 0.59 T for  $\gamma$-CoV$_2$O$_6$. In the  field range, $H_{c1}$$<$$H$$<$$H_{c2}$, $M$ is almost $H$ independent  and its value  is approximately 1/3 of the saturation magnetization.  Insets of Fig. 3 display the five-segment $M$($H$) curve.   $M$($H$) for both the samples shows a small hysteresis at low field. We observe that the hysteresis  decreases rapidly with increasing temperature.  Figs. 3(a) and (b) show that the sharpness of the metamagnetic transition reduces rapidly and the magnetization plateau  progressively weakens with increasing temperature and disappears above a critical temperature. For these aligned samples, the values of $M$ and critical fields as well as the nature of $M$($H$) curve at a given $T$  are very similar to that reported for single crystals. \cite{zhe,lenertz} However, the overall nature of $M$($H$) curve for the single crystal and aligned samples of CoV$_{2}$O$_{6}$ is very different from that of randomly oriented polycrystalline samples. For example,  $M$($H$) for polycrystalline sample does not show step-like increase, clear 1/3 magnetization plateau and any tendency of saturation. \cite{nandi} Neutron diffraction studies have revealed that CoV$_{2}$O$_{6}$ compounds undergo successive field-induced transitions from AFM to ferrimagnetic (FI) state at $H_{c1}$ and FI to FM state at  $H_{c2}$.\cite{lener,markk,lenertz}  At 2 K, the values of saturation magnetization are 4.5 and 2.9 $\mu_B$/Co ion, respectively for  $\alpha$ and $\gamma$ phases which are close to those reported for the single crystals. \cite{zhe,lenertz} Though the value of  saturation magnetization for the $\gamma$ phase is close to the expected spin-only moment (3.0 $\mu_B$/Co) of high-spin Co$^{2+}$, the huge difference between the observed saturation moment and expected spin-only moment in $\alpha$ phase indicates a strong spin-orbit coupling in this system.\cite{zhe} On the other hand, the effective moment derived from the Curie-Weiss plot is significantly larger than the spin only moment for both the phases. This kind of discrepancy in the derived value of magnetic moment  has also been observed in single crystalline samples. \cite{lenertz,dree}   It appears that simple magnetic measurements cannot resolve the relative contribution of spin and orbital magnetic moment in CoV$_{2}$O$_{6}$.  \\

The spin as well as orbital contribution to magnetic moment for both $\alpha$ and $\gamma$ phases have been determined from the X-ray magnetic circular dichroism (XMCD) experiment where the value of orbital and spin moments are found to be 1.9  and 2.5 $\mu_B$, respectively for the $\alpha$ phase and the corresponding values are 0.7  and 1.8 $\mu_B$ for the $\gamma$ phase\cite{holl}. Though the value of total magnetic moment deduced from  XMCD experiment is significantly smaller than the saturation magnetization for $\gamma$-CoV$_2$O$_6$, the above result clearly indicates an exceptionally high and a moderate orbital contribution to the magnetism in $\alpha$-CoV$_2$O$_6$ and $\gamma$-CoV$_2$O$_6$, respectively.  Magnetization measurements on single crystalline samples reveal  saturation moment  3.2 $\mu_B$/Co for $\gamma$-CoV$_2$O$_6$. \cite{dree} Furthermore,  the magnetic moment is found to increase slowly with magnetic field and it exceeds 3 $\mu_B$/Co above 7 T for magnetically aligned polycrystalline sample. \cite{lenertz} These observations clearly suggest the presence of spin-orbit coupling in $\gamma$-CoV$_2$O$_6$.  Recently, Wallington {\it et al}. have studied the spin-orbit transitions in both $\alpha$ and $\gamma$ polymorphs by neutron inelastic scattering.\cite{wall}  In monoclinic $\alpha$-CoV$_2$O$_6$, they observe that the spin exchange is  weak in comparison with the spin-orbit coupling. On the other hand, the spin exchange is larger as compared to spin-orbit coupling in $\gamma$-CoV$_2$O$_6$.  The significant difference in spin-orbit coupling between two polymorphs of CoV$_2$O$_6$ has been attributed to the Co-Co distance and the local deformation of the octahedral oxygen environment around the Co$^{2+}$ ion. \cite{lene,lenertz,wall}  CoO$_6$ octahedron is highly distorted in $\alpha$ phase whereas it is almost regular in $\gamma$ phase. Due to this large distortion of CoO$_6$ octahedron, the spin-orbit coupling is quite strong in $\alpha$  phase.  In order to understand the unusual magnetic properties of CoV$_2$O$_6$ system and to estimate the spin and orbital magnetic moment contribution several theoretical calculations have been done. \cite{kim,wall,sau,yao} Theoretical calculations also suggest that the orbital contribution to magnetic moment in $\alpha$-CoV$_2$O$_6$ is significantly larger than that in  $\gamma$-CoV$_2$O$_6$.\cite{kim} However, the large difference in the value of magnetic moment derived from ferromagnetic and paramagnetic states remains unanswered.\\

\begin{figure}
\begin{center}
\includegraphics[width=0.50\textwidth]{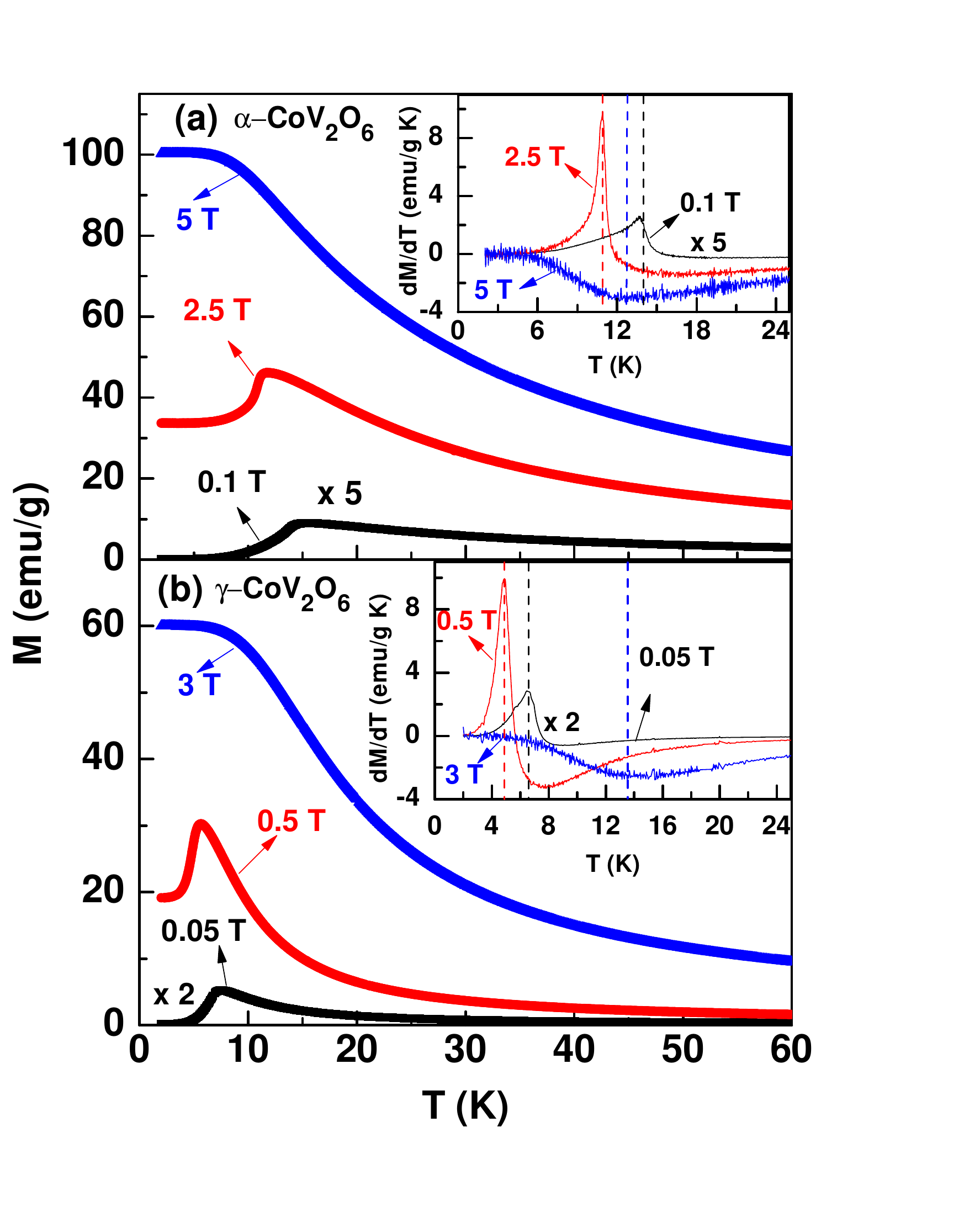}
\caption{(a) and (b) Temperature dependence of magnetization at three representative magnetic fields reflecting the three field-induced magnetic order states of $\alpha$-CoV$_{2}$O$_{6}$ and $\gamma$-CoV$_{2}$O$_{6}$, respectively. For clarity $M$($T$) at 0.1 T of $\alpha$-CoV$_{2}$O$_{6}$ is magnified 5 times while $M$($T$) at 0.05 T of $\gamma$-CoV$_{2}$O$_{6}$ is magnified 2 times. Insets show derivatives of $M$ versus $T$ curves (d$M$/d$T$) where vertical lines indicate transition temperatures.}
\end{center}
\end{figure}
For better understanding the nature of field-induced magnetic transitions, the temperature dependence of magnetization  at different fields both below and above the critical fields has  been measured.  The temperature dependence of magnetization is shown in Fig. 4 for three selected fields. Three different field-induced ordered magnetic states namely antiferromagnetic, ferrimagnetic and ferromagnetic are clearly reflected from these representative curves. For $\gamma$-CoV$_2$O$_6$, the peak  at 6.5 K in $M$($T$) curve due to the AFM-PM transition disappears for an applied field of 0.5 T  but a new sharp peak appears at a lower temperature around 4.8 K. As the applied field is in between  $H_{c1}$ and $H_{c2}$, the peak at 4.8 K is due to the FI-PM transition. Similar behavior is observed for the $M$($T$) curve of the $\alpha$-CoV$_{2}$O$_{6}$ sample at 2.5 T. In this case, the FI-PM transition occurs at $T_{FI}$$=$11 K.  For both the systems, initially  $M$ decreases slowly with decreasing $T$ below $T_{FI}$ and then starts to saturate at low temperature.   As expected,  $M$ in the ferrimagnetic state is about 1/3 of the saturation magnetization for both the samples. The temperature and magnetic field range over which the field-induced ferrimagnetic phase observed is much narrower in $\gamma$ phase.  For applied field above $H_{c2}$, $M$ increases monotonically with decreasing temperature and the nature of $M$($T$) curve  is very similar to that for a ferromagnetic system. However, the transition region is quite broad due to the high value of applied magnetic field. For $\alpha$-CoV$_{2}$O$_{6}$, both FI-PM  transition temperature (11 K at 2.5 T) and FM-PM transition  temperature (12.7 K at 5 T) are close to each others. On the other hand, the FM-PM transition for $\gamma$-CoV$_{2}$O$_{6}$  takes place around 13 K at 3 T which is well above $T_N$.  The FM-PM transition temperature ($T_{C}$) was determined from the position of the minimum in d$M$/d$T$ versus $T$ curve as shown in the insets of Fig. 4.  For  better clarification of the AFM-PM, FI-PM and FM-PM transitions, the derivatives of $M$($T$) curves are also presented in the insets of Fig. 4. Thus, $T_{C}$ is comparable for both the compounds though their AFM and FI transitions, critical fields and saturation values of magnetization are significantly different.  \\
\begin{figure}[b!]
\begin{center}
\includegraphics[width=0.50\textwidth]{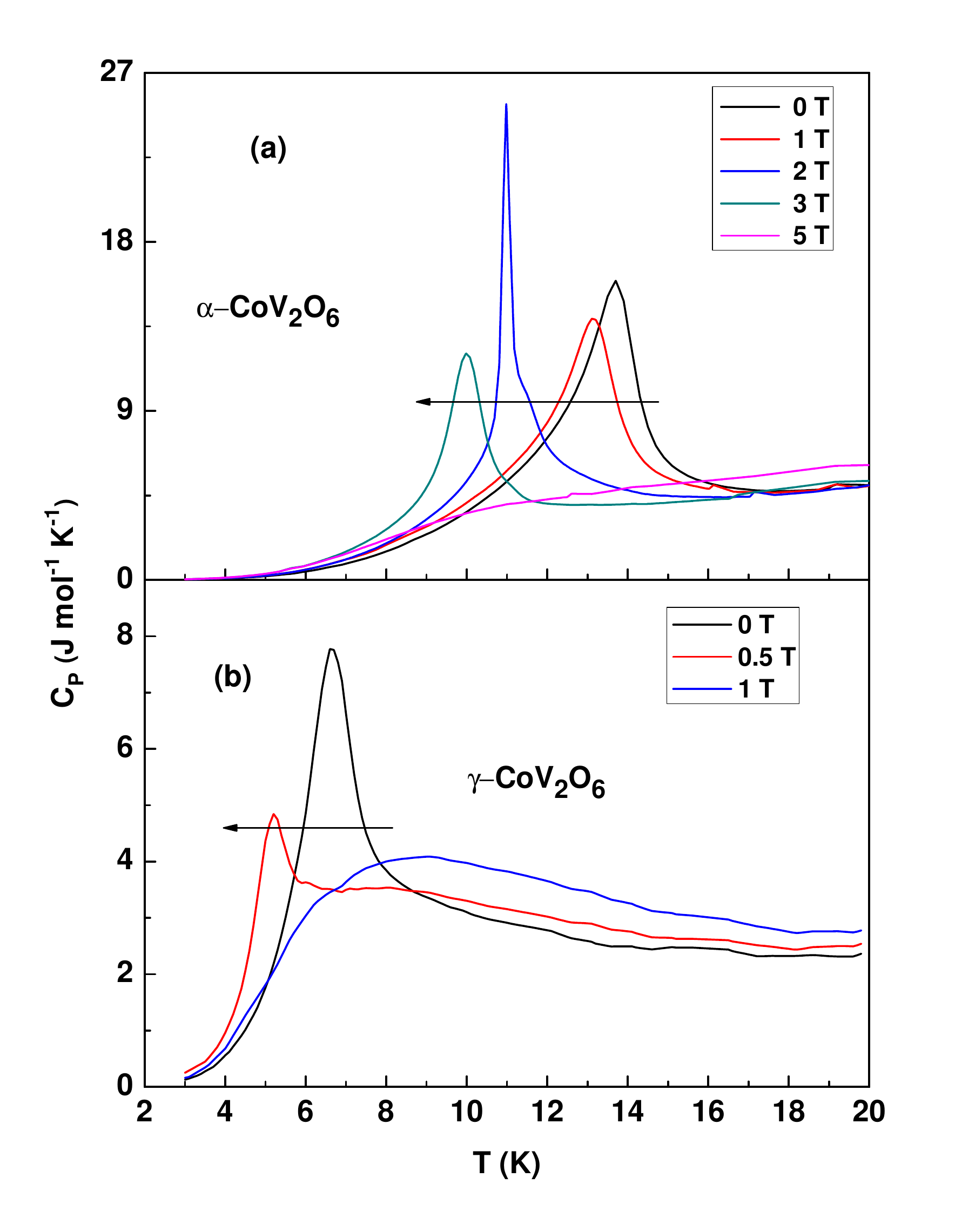}
\caption{(a) and (b) Temperature dependence of specific heat for aligned $\alpha$-CoV$_{2}$O$_{6}$ and $\gamma$-CoV$_{2}$O$_{6}$, respectively, for some selected fields. Arrow indicates the direction of increase of field.}
\end{center}
\end{figure}
Temperature dependence of the specific heat for the aligned $\alpha$-CoV$_{2}$O$_{6}$ and $\gamma$-CoV$_{2}$O$_{6}$ samples has been plotted in Figs. 5(a) and (b), respectively for some selected fields. For both the samples, the magnetic field was applied parallel to  the easy axis of magnetization. In absence of external magnetic field,  the specific heat curve shows a $\lambda$-like peak around $T_N$  similar to  that observed in polycrystalline random sample. With the application of magnetic field, the peak gradually shifts toward lower temperature and its height decreases but the nature does not change from $\lambda$-like  up to $H_{c1}$.  However, the nature of peak for $\alpha$-CoV$_{2}$O$_{6}$ changes abruptly above  $H_{c1}$. For example, the $\lambda$-like peak due to AFM transition disappears and a very sharp peak appears around 11 K at 2 T field ($H_{c1}$$<$2 T$<$$H_{c2}$).  The emergence of this new peak around $T_{FI}$ $=$11 K  is due to the transition from ferrimagnetic to paramagnetic state. One can see that the nature of this peak is completely different from the peak at $T_{N}$. The peak at $T_{FI}$ is extremely sharp and symmetric. The full-width at half-maximum for the peak at $T_{FI}$  is much smaller than that for the peak at $T_{N}$. At 3 T, the peak is much less shaper as compared to that for 2 T field. This is because the applied field is close to the critical field $H_{c2}$. The effect of magnetic field on magnetic transition is more clearly reflected in the derivative of $M$($T$) curve at different fields which are  presented in the inset of Fig. 4. From those plots, one can see that the peak at $T_{FI}$ is much sharper than the peak at $T_{N}$. It is worth mentioning that $\alpha$-CoV$_{2}$O$_{6}$ exhibits huge magnetostriction effect below $T_N$ and, similar to  heat capacity, the coefficient of thermal expansion  exhibits a sharp peak at $T_{FI}$ for an applied field of 2 T. \cite{nandi} This suggests that magneto-elastic coupling in $\alpha$-CoV$_{2}$O$_{6}$ is quite strong. Usually, the delta-like sharp and symmetric peak in heat capacity indicates first order nature of phase transition. But one cannot determine the order of a transition only by observing the nature of the peak in heat capacity. Thermal hysteresis is important for determining whether the transition is first order or not. We have not observed any thermal hysteresis either in  $M$($T$) or in $C_{p}$($T$) within the resolution of our measurement. This suggests that the transition is either second order or weakly first order in nature. At 5 T ($H$$>$$H_{c2}$), the specific heat curve exhibits a hump-like feature at the FM to PM transition. Unlike $\alpha$-CoV$_{2}$O$_{6}$, the peak in $C_{p}$($T$) of $\gamma$-CoV$_{2}$O$_{6}$ progressively suppresses with the increase of magnetic field strength. The sharpness of  $C_{p}$($T$) peak reduces appreciably in the ferrimagnetic state and the peak becomes very broad in the FM state. \\

Lenertz {\it et al}. first observed additional steps in $M$($H$) curve for $\alpha$-CoV$_{2}$O$_{6}$  at low temperature \cite{lene}. Similar to their report, we have also observed additional magnetization steps in $\alpha$-CoV$_{2}$O$_{6}$ in both increasing and decreasing field cycles. The additional magnetization steps are shown in Fig. 6 for some selected temperatures. For example,  the field-induced steps are observed at 1.6, 2.2, 2.8 and 3.4 T at 1.8 K. The  two additional field-induced transitions at $H_{cs1}$ $\sim$ 2.2 T and $H_{cs2}$ $\sim$ 2.8 T are quite prominent.  With the increase of temperature, the sharpness of these transitions reduces rapidly and only two field-induced transitions at  $H_{c1}$ (1.5 T) and  $H_{c2}$ (3.3 T) continue to survive  above 6 K. So, it is clear that the number of the magnetization steps is highly sensitive to temperature. However, the number of steps is independent of the sweep rate of the  magnetic field. At low temperature, magnetic field driven multiple steps are detected in spin chain Ca$_{3}$Co$_{2}$O$_{6}$ in a regular field interval and the number of steps in this system depends on the sweep rate of the external magnetic field and temperature\cite{maignan,hardy,maignan1}. The appearance of additional steps in Ca$_{3}$Co$_{2}$O$_{6}$ at a regular interval of applied magnetic field below 10 K was attributed to the emergence of metastable states and was explained by the Monte Carlo simulation with 2D triangular lattice of Ising chain \cite{kudasov,soto}. However,  the reason behind multiple magnetization steps in $\alpha$-CoV$_{2}$O$_{6}$ seems to be  different because these  steps appear in irregular field intervals. The Monte Carlo simulation also suggests that the multiple steps in this system appear due to the effect of different sublattices of 3D structure\cite{kim}.\\
\begin{figure}
\begin{center}
\includegraphics[width=0.50\textwidth]{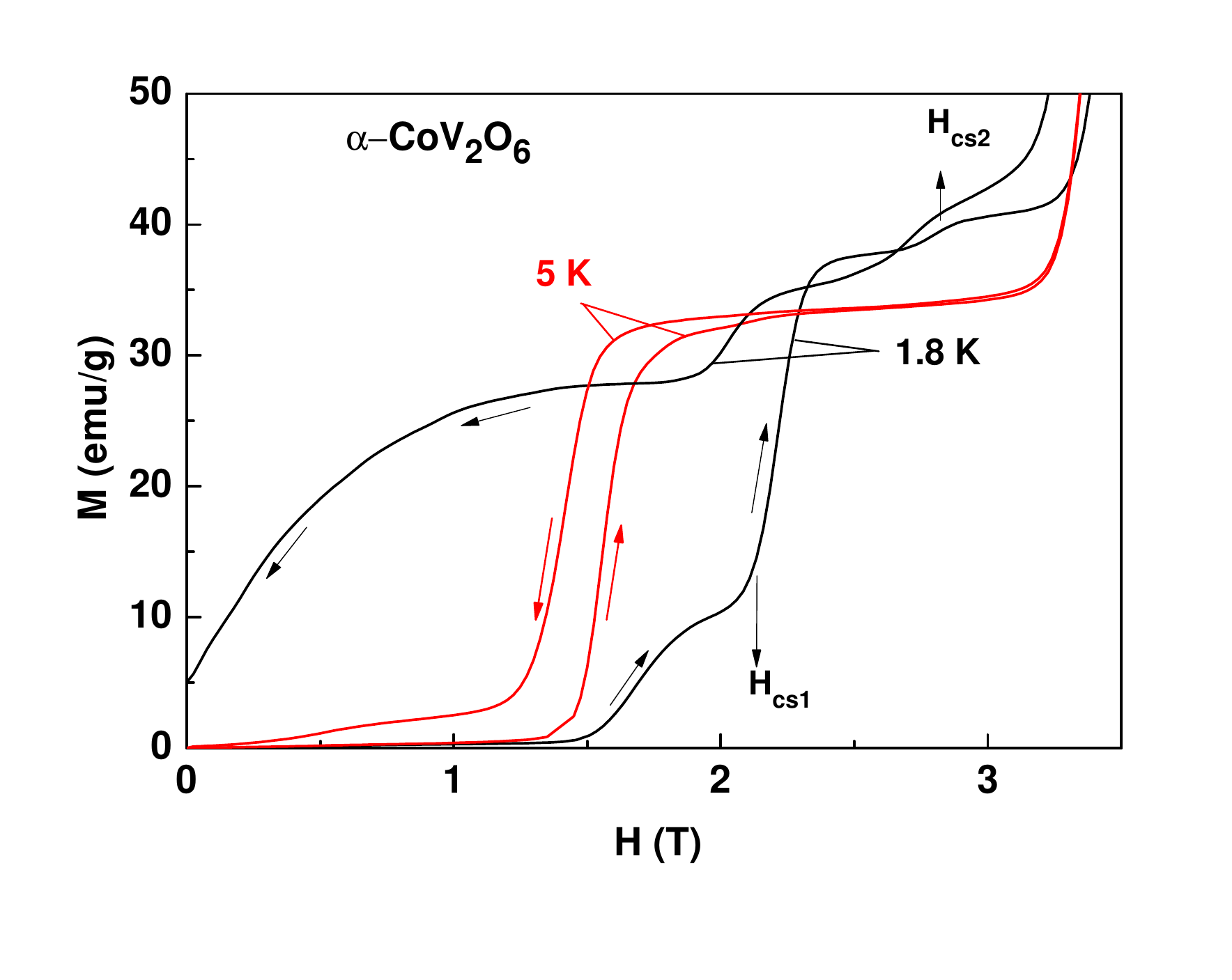}
\caption{Field dependence of magnetizations for $\alpha$-CoV$_{2}$O$_{6}$ for both increasing and decreasing field is plotted for some selected temperatures below 6 K.}
\end{center}
\end{figure}\\

\begin{figure}
\begin{center}
\includegraphics[width=0.50\textwidth]{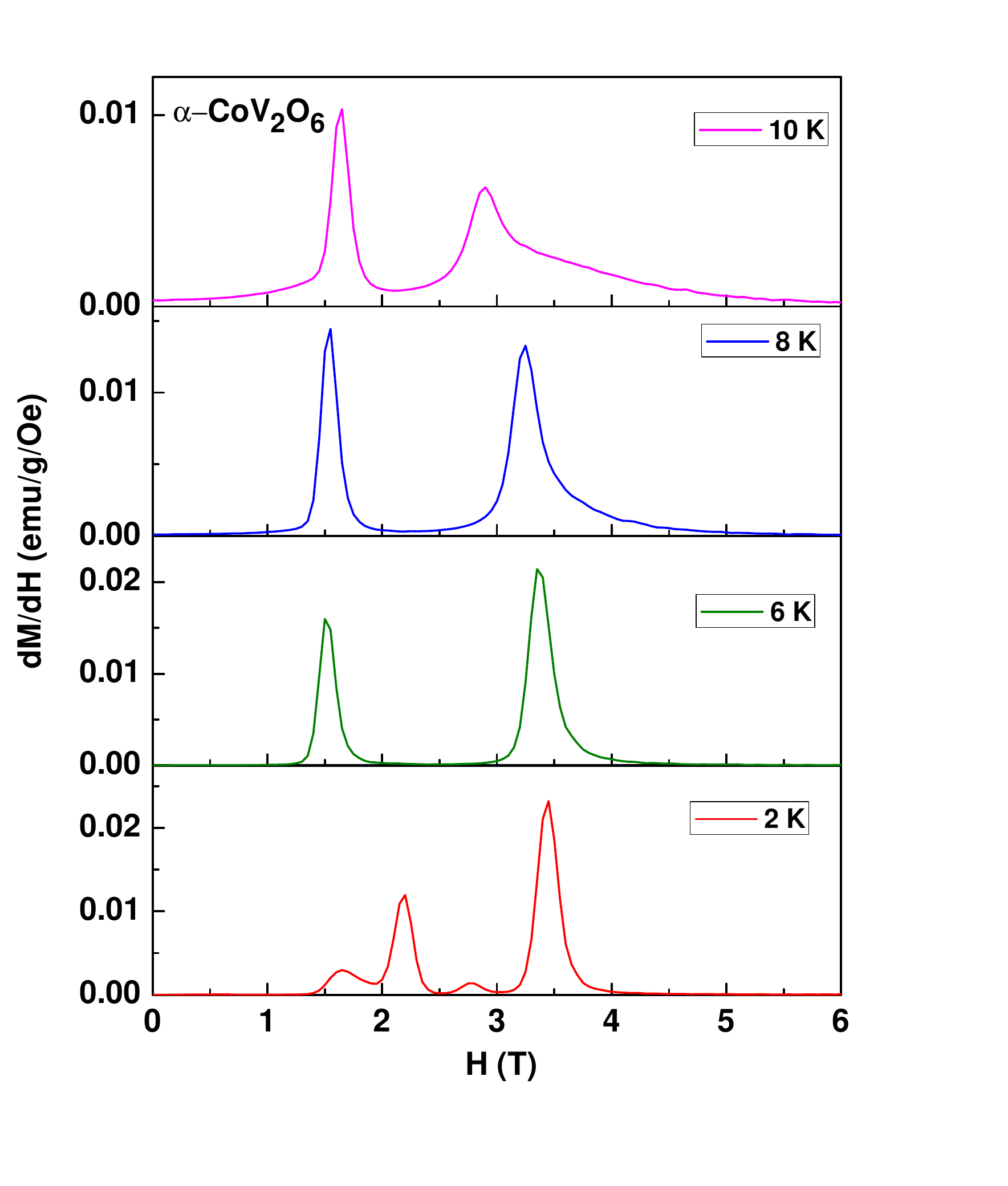}
\caption{Some selective derivatives of $M$($H$) curves (d$M$/d$H$) for $\alpha$-CoV$_{2}$O$_{6}$ are shown in the temperature range 2-10 K.}
\end{center}
\end{figure}
\begin{figure}
\begin{center}
\includegraphics[width=0.50\textwidth]{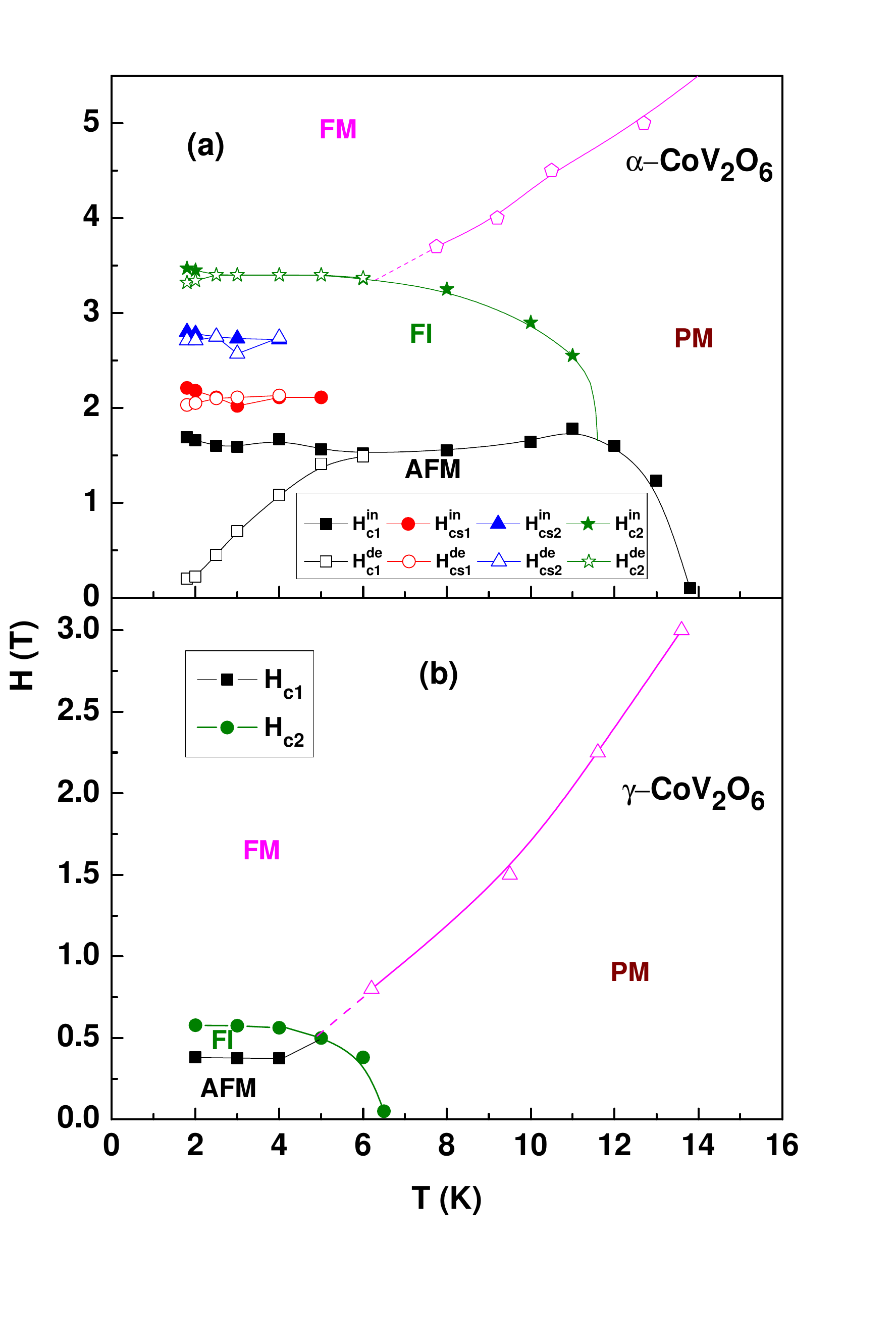}
\caption{(a) and (b) show magnetic phase diagram for $\alpha$-CoV$_{2}$O$_{6}$ and $\gamma$-CoV$_{2}$O$_{6}$, respectively. For $\alpha$-CoV$_{2}$O$_{6}$, the temperature dependence of four critical fields $H_{c1}$, $H_{cs1}$, $H_{cs2}$ and $H_{c2}$ for both field increasing ($H^{in}$) and decreasing ($H^{de}$) conditions is presented. For $\gamma$-CoV$_{2}$O$_{6}$, the temperature dependence of two critical fields $H_{c1}$ and $H_{c2}$ is presented. Antiferromagnetic (AFM), ferrimagnetic (FI), ferromagnetic (FM) and paramagnetic (PM) states are also depicted in these graph.}
\end{center}
\end{figure}
For the detailed investigation of the nature of the metamagnetic transition in $\alpha$-CoV$_{2}$O$_{6}$, we have calculated the derivatives of $M$($H$) curves in the temperature range 1.8 K to 20 K. A few representative d$M$/d$H$ versus $H$ curves are shown in Fig. 7 for some selected temperatures where four critical fields $H_{c1}$, $H_{cs1}$, $H_{cs2}$, $H_{c2}$ have been identified. Similarly the critical temperatures were determined from $M$($T$) curves. The critical temperatures and fields determined from these  curves are plotted as a function of temperature  in Fig. 8(a). $H_{c1}$ exhibits a hysteresis below 6 K which increases with decreasing temperature and becomes huge at 1.8 K. Above 6 K, $H_{c1}$ does not show temperature dependence up to 10 K and then increases slowly, exhibits a broad maximum  around 11 K and  disappears just below 14 K. Both $H_{cs1}$ and $H_{cs2}$ show very weak temperature dependence. $H_{c2}$ also remains almost constant up to 6 K and then decreases rapidly with further increase of temperature and disappears above 11 K. A small hysteresis has also been observed in $H_{c2}$ below 2.5 K. Another interesting point to be mentioned here is the  nature of peaks around $H_{c1}$ and $H_{c2}$ in Fig. 7. Both the peaks are highly symmetric up to 6 K but the peak at $H_{c2}$ becomes slightly asymmetric just above 6 K. With increase of temperature above 6 K, the asymmetric nature increases rapidly,  the peak becomes very broad and shifts towards lower field. However, the peak around $H_{c1}$ remains symmetric.   Finally, three  characteristic temperatures $T_{c1}$, $T_{c2}$ and $T_{c3}$  are identified around 6, 11 and 14 K, respectively from $T$ and $H$ dependence of $M$  and d$M$/d$H$ versus $H$ graphs. $T_{c2}$ and $T_{c3}$ are identified as $T_{FI}$ and $T_{N}$, respectively. Below $T_{c1}$, we observe several significant changes in the nature of critical field and magnetization such as the appearance of hysteresis in $H_{c1}$ and steps at $H_{cs1}$ and $H_{cs2}$ in $M$. Also, a sharp increase in magnetic entropy  is observed just below $T_{c1}$ (to be discussed in the later section). However, heat capacity measurements do not show any peak or anomaly around $T_{c1}$. Thus we cannot say that $T_{c1}$ is some critical temperature associated with thermodynamic phase transition. The temperature and field dependence of $M$ and $C_{p}$ have also been analyzed to draw the phase diagram for  $\gamma$-CoV$_{2}$O$_{6}$ system [Fig. 8(b)]. For both the systems, the phase boundary between FM and PM state has  been determined  from the $dM/dT$ versus $T$ curves for different applied fields.  Unlike $\alpha$-CoV$_{2}$O$_{6}$, the phase diagram is quite simple for  $\gamma$-CoV$_{2}$O$_{6}$.   Kimber {\it et al}. reported the ($H$-$T$) phase diagram for $\gamma$-CoV$_{2}$O$_{6}$ system based on magnetization and neutron diffraction data of polycrystalline random sample. \cite{kimb} According to their phase diagram, there exists a significant portion  where the system is FM well below $H_{c2}$. For example, in the temperature region 5 K$<$$T$$<$7 K, $\gamma$-CoV$_{2}$O$_{6}$ is FM for an applied field of 0.35 T which is well below the critical field 0.59 T for the FM transition of that sample. Furthermore, the FM-PM transition temperature is observed to decrease rapidly with increasing magnetic field. If this behavior continues then $\gamma$-CoV$_{2}$O$_{6}$ will become PM at low temperature for an applied  field of  3 T only which is contrary to experimental observation. \\

\begin{figure}[b!]
\begin{center}
\includegraphics[width=0.50\textwidth]{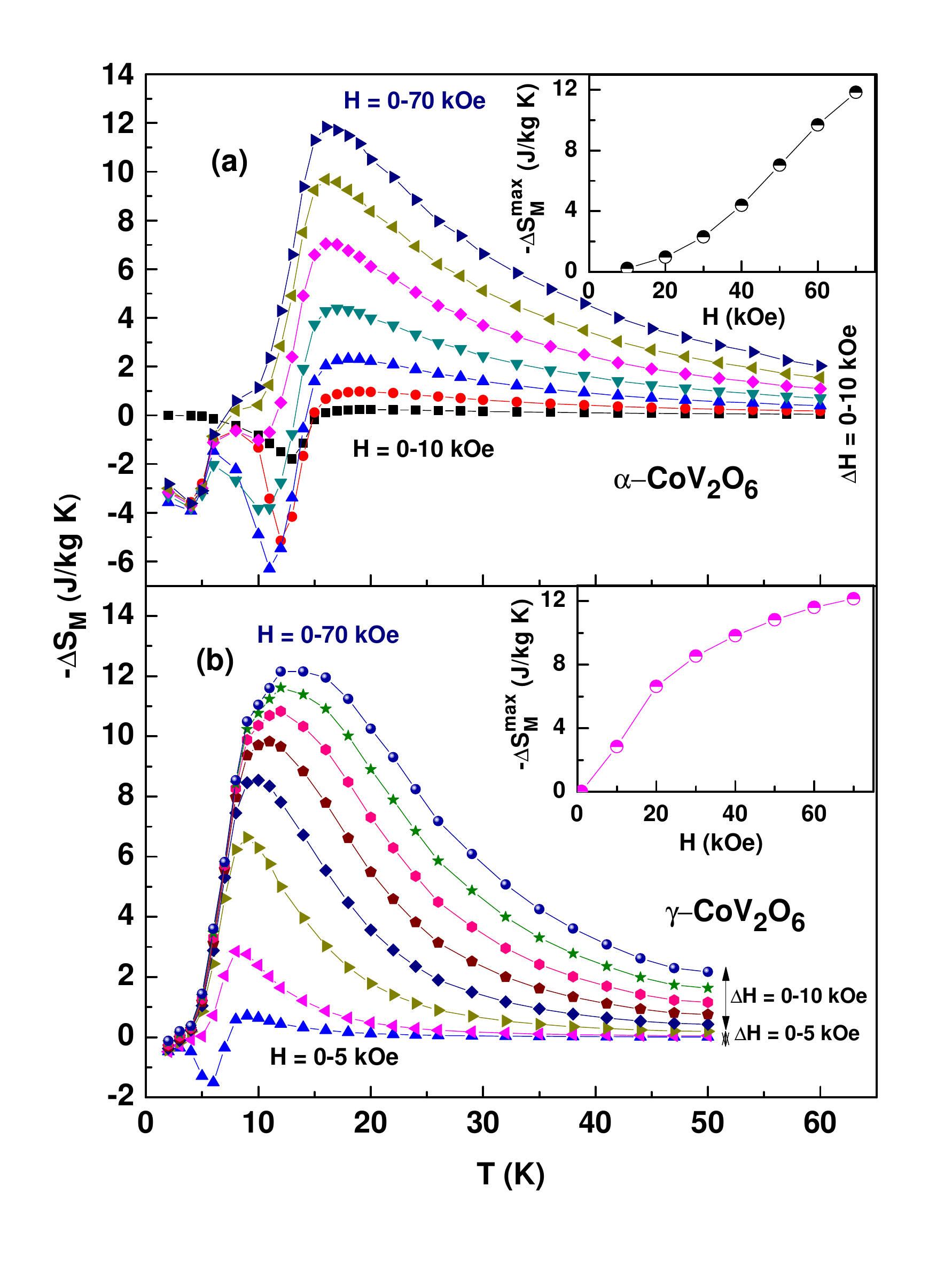}
\caption{Temperature dependence of magnetic entropy change -$\Delta S_{M}$ for  $\alpha$-CoV$_{2}$O$_{6}$ (a) and  $\gamma$-CoV$_{2}$O$_{6}$ (b). Insets show -$\Delta$S$_{M}^{max}$ as a function of magnetic field for $\alpha$-CoV$_{2}$O$_{6}$  and  $\gamma$-CoV$_{2}$O$_{6}$.}
\end{center}
\end{figure}
Often the field-induced metamagnetic transition causes huge magnetic entropy change. The isothermal  entropy change  can be derived from the well known Maxwell relation $\Delta S_{M}(T,H)=\int_{0}^{H}(\frac{\partial M}{\partial T})_{H} dH$. To calculate the magnetic  entropy change from the experimental data on magnetization,  the integral  equation  can be written in a summation form
\begin{equation}
\Delta S_{\rm M} = \sum_{i}\frac{M_{i+1} - M_{i}}{T_{i+1} -T_i} \Delta H_i
\label{eq1}
\end{equation}
where $M_{i}$ and $M_{i+1}$ are the measured magnetization at temperatures $T_{\rm i}$ and $T_{\rm i+1}$, respectively for a small change in magnetic field $\Delta${$H_{\rm i}$. $\Delta$S$_{M}$ can be deduced using the $M$($H$) curves of Fig.  3 at different temperatures. The magnetic entropy change for some selected magnetic fields is shown  in Figs. 9(a) and (b) as a function of temperature for $\alpha$-CoV$_{2}$O$_{6}$ and $\gamma$-CoV$_{2}$O$_{6}$, respectively. The temperature dependence of  entropy change for $\alpha$-CoV$_{2}$O$_{6}$ is quite complex due to the multiple field-induced magnetic transitions. For both the systems, the entropy change can be negative or positive depending on field and temperature. In other words, these systems exhibit both conventional (negative $\Delta$S$_{M}$) and inverse (positive $\Delta$S$_{M}$) MCE. \\

In the PM state, -$\Delta$S$_{M}$ increases rapidly with decreasing temperature and increasing field due to the suppression spin-disordering by external magnetic field. For $H$$\ll$$H_{c1}$, -$\Delta$S$_{M}$ starts to decrease  as $T$ approaches towards $T_N$ owing to suppression of sublattice magnetization in the AFM state. This behavior of  -$\Delta$S$_{M}$($T$) reveals a broad maximum (conventional MCE) slightly above $T_N$  and a minimum (inverse MCE) at $T_N$ for both the systems. The minimum in -$\Delta$S$_{M}$($T$) shifts slowly towards lower temperature with the increase of magnetic field and  reaches  a  value as low as -6.2 J kg$^{-1}$ K$^{-1}$ at 3 T for the $\alpha$ phase and -1.5 J kg$^{-1}$ K$^{-1}$ at 0.5 T for the $\gamma$ phase.  These minima occur at $T_{FI}$ for applied field below $H_{c2}$. Thus, the ferrimagnetic phase exhibits large inverse MCE. -$\Delta$S$_{M}$($T$) also shows minimum at 4 T which is slightly higher than $H_{c2}$. It is clear  that  the position of maximum in -$\Delta$S$_{M}$($T$) for $\alpha$-CoV$_{2}$O$_{6}$ does not shift significantly in the measured field range. The maximum  shifts slowly towards lower temperature from 19 to 16 K as magnetic field increases from 1 to 5 T and no shift is observed above  5 T. On the other hand, the maximum in -$\Delta$S$_{M}$($T$) for $\gamma$-CoV$_{2}$O$_{6}$ shifts progressively towards higher temperature from  9 K ($\sim$$T_N$) to 14 K ($\sim$2$T_N$) as field increases from 0.5 to 7 T. This difference in the nature of temperature dependence of magnetic entropy change can be explained on the basis of relative strength of AFM interaction in these two systems.  Unlike $\alpha$-CoV$_{2}$O$_{6}$, $\gamma$-CoV$_{2}$O$_{6}$ becomes FM with $T_C$ higher than $T_N$ at a relatively small applied field due to its weaker AFM interaction and $T_C$ increases rapidly with increasing field strength.  For this reason, the  maximum in -$\Delta$S$_{M}$($T$) for $\gamma$-CoV$_{2}$O$_{6}$ occurs close to $T_C$ and shifts towards higher temperature side when the applied field exceeds $H_{c2}$. Several features related to the field induced-magnetic transitions are  clearly reflected in -$\Delta$S$_{M}$($T$) curve. The most interesting observation is the step-like decrease in -$\Delta$S$_{M}$  around 6 K for fields above 1 T which coincides with $T_{c1}$. Thus, the appearance of field-induced additional magnetization steps and the step-like decrease in magnetic entropy change below a certain temperature indicate the existence of complex magnetic phases in the low-temperature region in $\alpha$-CoV$_{2}$O$_{6}$.\\

The inverse MCE is usually observed when a system undergoes field-induced first-order magnetic transformations such as from AFM to FM in  Fe$_{0.49}$Rh$_{0.51}$\cite{nik}, collinear-AFM to non-collinear-AFM in Mn$_5$Si$_3$\cite{teg} or AFM to FI in Mn$_{1.82}$V$_{0.18}$Sb.\cite{zha} Inverse MCE  is also observed in several rare-earth transition metal based antiferromagnetic compounds.\cite{kag,hu,naik,mid1,jin,mid2} When an external magnetic field is applied along the easy axis, the magnetic moment fluctuation is enhanced in one of the two AFM sublattices which is antiparallel to $H$. With the increase of $H$, more and more spins in the antiparallel sublattice orient along the field direction. This, in turn, increases the spin disordering  and, hence inverse MCE occurs.  Usually, such kind of behavior continues up to a certain field.  As the majority of spins in the antiparallel sublattice orient along the field direction at high fields, the system is expected to show conventional MCE. In other wards, the decrease in magnetic entropy upon the application of magnetic field is associated with FM alignment of spins as the Zeeman energy suppresses spin fluctuations. With the increase of field, the value of -$\Delta$S$_{M}$ at peak (-$\Delta$S$_{M}^{max}$) increases rapidly for both the systems.  Insets of Figs. 9(b) and (b) show the field dependence of -$\Delta$S$_{M}^{max}$.  For $\gamma$-CoV$_{2}$O$_{6}$, initially -$\Delta$S$_{M}^{max}$ increases at a faster rate with field and tends to saturate at high fields. As a result, this system exhibits large entropy change at a relatively small applied field.   On the other hand, -$\Delta$S$_{M}^{max}$  for $\alpha$-CoV$_{2}$O$_{6}$ increases at a much slower rate in the low-field region and does not show any sign of saturation up to 7 T. For both the systems, the value of -$\Delta$S$_{M}^{max}$  is  about 12 J kg$^{-1}$ K$^{-1}$ at 7 T. In the low-field region, the faster increase of -$\Delta$S$_{M}^{max}$ with field in $\gamma$-CoV$_{2}$O$_{6}$ as compared to $\alpha$-CoV$_{2}$O$_{6}$  is due to the smaller $H_{c2}$  of the former system. \\
\begin{figure}[b!]
\begin{center}
\includegraphics[width=0.50\textwidth]{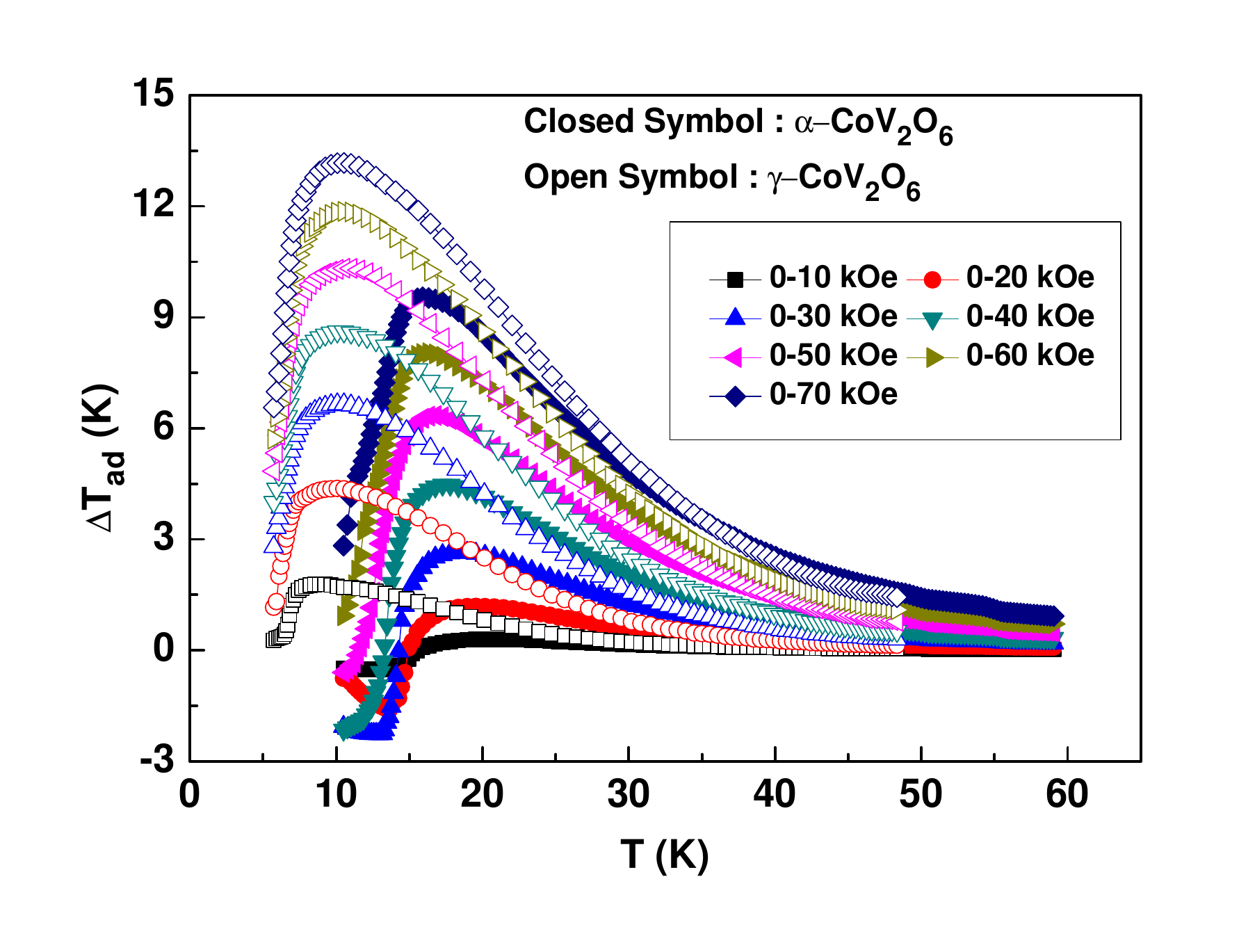}
\caption{ Temperature dependence of adiabatic temperature change ($\Delta T_{ad}$) for $\alpha$-CoV$_{2}$O$_{6}$ (closed symbol) and $\gamma$-CoV$_{2}$O$_{6}$ (open symbol).}
\end{center}
\end{figure}
We have also calculated another very important parameter related to magnetic refrigeration,  adiabatic temperature change, from the field-dependent magnetization and zero-field heat capacity data. The isentropic temperature difference between $S(0,T)$ and $S(H,T)$ is known as $\Delta T_{ad}$, where the total entropy at zero field $S(0,T)$ can be evaluated from $S(0,T)$$=$$\int\limits_{0}^{T}\frac{C_p(0, T)}{T}dT$. In order to get $S(H,T)$, the corresponding $\Delta S_M$($H$) has been subtracted from $S(0,T)$. For both the systems, the temperature dependence of $\Delta T_{ad}$ for various magnetic field changes is shown in Fig. 10.  The maximum values of adiabatic temperature change $\Delta T_{ad}^{max}$ reach 9.5 K and 13.1 K for a field change of 7 T for $\alpha$-CoV$_{2}$O$_{6}$ and $\gamma$-CoV$_{2}$O$_{6}$, respectively. Thus, spin-chain CoV$_{2}$O$_{6}$ exhibits large adiabatic temperature change.  The application potential of a material as  a magnetic refrigerant depends on the values of   $\Delta$S$_{M}$ and $\Delta T_{ad}$ parameters at low or moderate fields. Similar to  -$\Delta$S$_{M}$,  $\Delta T_{ad}$ is also large for $\gamma$-CoV$_{2}$O$_{6}$ in the low-field region. For example,  the maximum values of -$\Delta$S$_{M}$ and $\Delta T_{ad}$ at 3 T are as high as 8.5 J kg$^{-1}$ K$^{-1}$ and 7 K, respectively for the $\gamma$-CoV$_{2}$O$_{6}$ while the corresponding values are about 2.5 J kg$^{-1}$ K$^{-1}$ and 3 K  for the $\alpha$-CoV$_{2}$O$_{6}$. This suggests that $\gamma$-CoV$_{2}$O$_{6}$ may be considered as a low temperature magnetic refrigerant for magnetocaloric cooling.  \\

The nature of  -$\Delta$S$_{M}$($T$) curve of CoV$_{2}$O$_{6}$  has several similarities and differences with that of the Ising spin chain system Ca$_3$Co$_2$O$_6$, which also shows 1/3 magnetization plateau. In Ca$_3$Co$_2$O$_6$ single crystal,  -$\Delta$S$_{M}$($T$) shows a broad maximum for applied fields above 3 T. However, this maximum occurs at $T_C$ and its position does not shift with field and the value of -$\Delta$S$_{M}^{max}$ is much smaller than that for the $\gamma$-CoV$_{2}$O$_{6}$ at low as well as high fields. For example, the values of -$\Delta$S$_{M}^{max}$ are about 1.5 and 6.5 J kg$^{-1}$ K$^{-1}$ for field change of 3 and 7 T, respectively. Similar to $\alpha$-CoV$_{2}$O$_{6}$, -$\Delta$S$_{M}$($T$) of Ca$_3$Co$_2$O$_6$  shows a minimum at low temperature in the FM state.  This behavior has been attributed to the short-range FM ordering. In $\alpha$-CoV$_{2}$O$_{6}$, the origin of the minimum at 4 T could be due to the short-range FM ordering similar to that in Ca$_3$Co$_2$O$_6$. Ca$_3$Co$_2$O$_6$  also displays inverse MCE well below $T_C$ but its value is significantly larger than that for CoV$_{2}$O$_{6}$.  Magnetic entropy change has also been reported for several other compounds such as Nd$_6$Co$_{1.67}$Si$_3$  with -$\Delta$S$_{M}^{max}$$\sim$5.5 J kg$^{-1}$ K$^{-1}$ for  $\Delta H$= 5 T and (Mn$_{0.83}$Fe$_{0.17}$)$_{3.25}$Ge with -$\Delta$S$_{M}^{max}$$\sim$6 J kg$^{-1}$ K$^{-1}$ for  $\Delta H$= 5 T which show step-like behavior in magnetization. \cite{hald,jdu} In both the cases, -$\Delta$S$_{M}^{max}$ is smaller than that for CoV$_{2}$O$_{6}$ compounds. The values -$\Delta$S$_{M}^{max}$ for these compounds are also quite small at low or moderate field change.\\

\section{Summary}

In summary, a clear step-like increase in $M$($H$) with 1/3 magnetization plateau has been observed in magnetically aligned $\alpha$-CoV$_{2}$O$_{6}$ and $\gamma$-CoV$_{2}$O$_{6}$. Both the systems show field-induced AFM to FI to FM transitions with increasing field. $T_{FI}$ and $T_C$ are almost same but smaller than $T_N$ in $\alpha$-CoV$_{2}$O$_{6}$ whereas $T_C$ is significantly larger than both $T_N$ and $T_{FI}$ in $\gamma$-CoV$_{2}$O$_{6}$. In spite of several differences, $T_C$ is comparable in two systems. Temperature dependence of heat capacity and magnetization show that the nature of transitions around $T_N$ and $T_{FI}$ is different in $\alpha$ phase. In $\alpha$-CoV$_{2}$O$_{6}$,  the low temperature magnetization reveals field-induced complex magnetic phases and  multiple magnetization plateaux below $T_{c1}$ $\sim$ 6 K. Both inverse and large conventional magnetocaloric effect have been observed in these compounds. Even at low field, the magnetocaloric parameters for $\gamma$-CoV$_{2}$O$_{6}$ are reasonably large. Based on the present data, we have constructed $H$-$T$ phase diagram for both the systems.    \\

\end{document}